\documentclass[%
 reprint,
 prx,
superscriptaddress,
 amsmath,amssymb,
 aps,
floatfix,
]{revtex4-2}

\usepackage{amsmath}
\usepackage{amsfonts}
\usepackage{amssymb}

\usepackage{placeins}
\usepackage{hyperref}%
\usepackage[english]{babel}
\usepackage[T1]{fontenc}
\usepackage{lmodern}
\usepackage{svg}
\usepackage[utf8]{inputenc}
\usepackage[scaled=0.95]{helvet}
\usepackage{color}
\usepackage[normalem]{ulem}

\usepackage{xcolor,calc}
\hypersetup{linkcolor = black, citecolor = black, urlcolor = blue}

\providecommand{\U}[1]{\protect\rule{.1in}{.1in}}

\definecolor{darkgreen}{RGB}{2, 153, 0} 
\definecolor{darkyellow}{RGB}{204, 153, 0} 
\newenvironment{subfigure*}{}

\hypersetup{
colorlinks=true,
linkcolor=blue,
}

\usepackage{graphicx}
\usepackage{dcolumn}
\usepackage{bm}

\begin{document}

\preprint{APS/123-QED}

\title{Interfacial orbital torques excite nanoscale terahertz magnons}
\author{Harshita Devda}

 \email{harshita.devda@uni-konstanz.de}
 \affiliation{Department of Physics, University of Konstanz, DE-78457 Konstanz, Germany}

\author{Peter M. Oppeneer}
\affiliation{Department of Physics and Astronomy, Uppsala University, P. O. Box 516, S-751 20 Uppsala, Sweden}

\author{Ulrich Nowak}
\affiliation{Fachbereich Physik, Universität Konstanz, DE-78457 Konstanz, Germany}

\date{\today}

\begin{abstract}
Exchange-dominated magnons in nanometer-thick ferromagnets extend to the terahertz regime through thickness quantization of perpendicular standing spin-wave (PSSW) modes. While interfacial spin–orbit torques (SOTs) have been shown to enable the excitation of such modes, the microscopic origin of 
the interfacial driving torque remains unclear. In particular, the coexistence of spin and orbital currents complicates the understanding. Here, we develop {and} use an atomistic framework that explicitly resolves interfacial symmetries and separates spin and orbital torque contributions. Exploiting a trilayer geometry for a thin ferromagnet sandwiched between non-magnetic layers, where the symmetry-controlled polarity of the interfacial torque produces mode-selective {magnon} excitation as observed in {the} recent experiment {of} Salikhov \textit{et al.}, Nature Phys. \textbf{19}, 529 (2023), we 
disentangle the different interfacial torque contributions. By decomposing the torque into magnetization-even (field-like) and magnetization-odd components, we identify the field-like torque as the dominant contribution responsible for the excitation. Crucially, isolating orbital and spin contributions reveals that the interfacial orbital torque provides the primary channel for the efficient excitation of exchange-dominated THz magnons in thin ferromagnets. Our results establish a microscopic basis for symmetry-engineered control of confined terahertz spin dynamics in magnetic multilayers.

\end{abstract}

\maketitle

\section{Introduction}

The excitation of exchange‐dominated terahertz (THz) magnons in nanometer-thick metallic ferromagnets can provide a promising pathway toward ultrafast spintronic technologies. In confined ferromagnetic films, higher-order perpendicular standing spin waves (PSSWs) possess nanometer wavelengths and multi-THz frequencies \cite{Kirschner2017, direct_zakeri_2013, Razdolski2017,Zakeri2021}, making their efficient and selective excitation challenging. Since uniform magnetic fields couple predominantly to long-wavelength modes, accessing short-wavelength exchange modes requires strongly localized interfacial driving mechanisms \cite{Razdolski2017,Ritzmann2020}.

As demonstrated in a recent experiment by Salikhov \textit{et al.}\ \cite{Salikhov2023} spin-orbit torques (SOTs) \cite{manchon2019} in a non-magnet\,(NM)/ferromagnet\,(FM)/non-magnet trilayer can provide precisely such a mechanism. Owing to their interfacial localization and ultrafast temporal response \cite{Guimaraes2020, david_2021}, these torques are particularly well suited for driving exchange-dominated PSSW modes under ultrafast pulse excitation. However, the microscopic origin of the dominant driving torque in this regime is not clearly understood, as the conventional interpretations emphasize spin-current–induced torques arising from spin Hall effect (SHE) \cite{Hirsch1999, Sinova2015} and spin Rashba-Edelstein effect (SREE)\cite{Edelstein1990, Dyakonov1971,Haney2013a, Haney2013b, Freimuth2014, Emori2016} generated spin accumulation. It is now firmly established that the presence of charge currents in certain metallic non-magnets also generates substantial orbital angular momentum accumulations via orbital Hall (OHE) \cite{Tanaka2008, Kontani2009, Dongwook2018, Dongwook2020,Salemi2022OHE} and orbital Rashba-Edelstein (OREE) \cite{Salemi2019} mechanisms. These orbital accumulations transfer angular momentum across interfaces and contribute to large, field-like and damping-like torques  on the ferromagnetic spins.

The coexistence of interfacially localized spin and orbital torque contributions, along-with their magnetization dependent counterparts \cite{Chuang2020, salemi2021, Salemi2022}, makes it challenging to identify the dominant mechanism responsible for THz magnon excitation, particularly in the ultrafast regime.
\begin{figure*}
    \centering
    \includegraphics[width=0.9\textwidth]{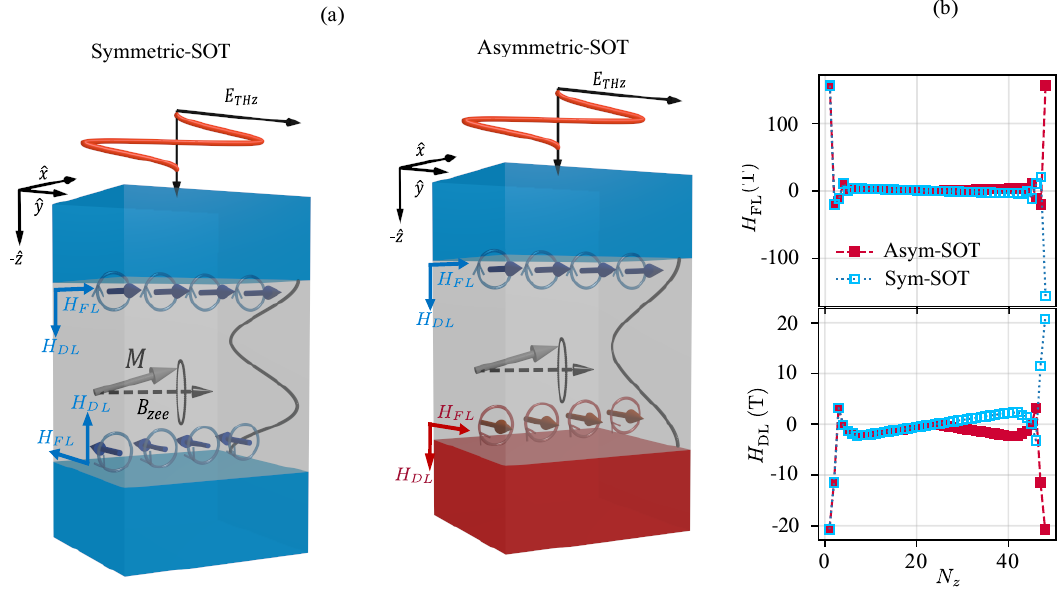}
   
    \caption{(a) Schematic illustration of symmetric (left) and asymmetric (right) trilayers, to which single cycle THz laser pulses are applied, generating  conventional SOTs at the two opposing interfaces. SOTs in symmetric/antisymmetric interfaces support odd/even PSSW modes. (b) Layer-resolved field-like and anti-damping-like SOT fields, $H_{\rm FL}$ and $H_{\rm ADL}$, respectively, taken from Ref.\ {\cite{Devda2025}} and extrapolated for $N_z=48$ {Co atomic layers,} for symmetric and asymmetric  interface configurations. 
    }
    \label{fig:trilayer_schematics}
\end{figure*}
Previous studies have shown that the interfacial orbital torque can exceed the spin torque \cite{Lee2021,Devda2025,Ding2020,Hayashi2023,Kim2021} in several respects
and also revealed that orbital torques are only weakly magnetization dependent, whereas spin torques exhibit a pronounced magnetization dependence \cite{Devda2025,Mahfouzi2018,Belashchenko2019}.
This distinction becomes especially relevant at THz frequencies, where rapid magnetization dynamics can strongly influence magnetization-dependent torque components \cite{Seifert2023,Choi2023,Kampfrath2013}. While the substantial magnitude of orbital torque suggests it may dominate the excitation mechanism, additionally, the strong magnetization dependence of spin accumulation implies that its contribution cannot be neglected under intense THz excitation \cite{Jungfleisch2018,Seifert2016}.

To investigate this, we combine here first-principles-calculated electrically induced spin and orbital moment accumulations for a thin Co/Pt system with atomistic spin-dynamics simulations to disentangle the respective roles of spin and orbital torque in the excitation of exchange-dominated PSSW modes in asymmetric and symmetric interfaced NM/Co/NM trilayers, {see Fig.\ \ref{fig:trilayer_schematics}(a)}. By incorporating the layer-resolved spin and orbital torques directly into the atomistic model, we demonstrate that high-frequency THz PSSW excitation is predominantly driven by interfacial orbital torques, while spin-current–induced torques remain comparatively inefficient for short-wavelength modes. Our calculated mode spectra and frequency selectivity are consistent with quantized exchange-dominated eigenmodes and quantitatively reproduce the experimentally observed behavior {\cite{Salikhov2023}}. Our results identify fast orbital angular momentum accumulation and corresponding torque as the primary mechanism enabling efficient excitation of THz exchange magnons in metallic heterostructures. 

\section{Atomistic spin Modeling of THz-SOT}

We start {with} modeling the trilayers sketched in Fig. \ref{fig:trilayer_schematics} {(a)} by defining the Hamiltonian to describe the spin interactions for a Pt/Co/Pt system. To model the ferromagnetic Co layer with atoms at position $i$ with strong spin magnetic moments, $\boldsymbol{\mu}_{S,i}=\mu^d_{S,i} \boldsymbol{S}_{i}$ and $|\boldsymbol{S}_{i}|=1$, {and} incorporating the interactions with the proximity induced moments in  the non-magnetic Pt layers, we utilized the renormalized Heisenberg Hamiltonian of Ref.\ \cite{Devda2025}, given as 
\begin{equation}
     \mathcal{H}^\text{eff}=  \sum_{i} \boldsymbol{S}_{i} {\widetilde{\mathcal{K}}_{i}} \boldsymbol{S}_{i} -\frac{1}{2}\sum_{i\neq j} \boldsymbol{S}_{i}\widetilde{\mathcal{J}}_{ij}\boldsymbol{S}_{j} - \sum_i \mu_{i,s} \boldsymbol{B}_i \cdot \boldsymbol{S}_i \, ,
     \label{eq:Ham_renorm-zero-field} 
\end{equation}
with renormalized exchange tensors $\widetilde{\mathcal{J}}_{ij}$ 
and renormalized on-site anisotropy tensors $\widetilde{\mathcal{K}}_{i}$, along with an additional Zeeman term for the magnetic field $\boldsymbol{B}$. The latter is of magnitude  $400$ mT and is applied in-plane orienting the ground state magnetization in $\hat{y}$ direction. 

The SOT field generated in the presence of an applied electric THz field is incorporated as extended Heisenberg Hamiltonian  with 
an additional interaction term $\mathcal{H}^{\rm{SOT}}$ as (see \cite{Devda2025} for further details)
\begin{align}
     \mathcal{H} & = \mathcal{H}^\text{eff} +\mathcal{H}^{\rm{SOT}} ,
\label{eq:ham-renorm-with-field_kurz}
    \end{align}
    where, 
\begin{align}
   & \mathcal{H}^\text{SOT}=  -\sum_{i\neq j} \boldsymbol{S}_{i} \mathcal{J}_{ij} \frac{\delta\bm{\mu}_{S,j}}{\mu_{S,i}^{d}}-\sum_{i} J^{sd} \, \frac{\delta\bm{\mu}_{S,i}}{\mu_{S,i}^{d}} \cdot \boldsymbol{S}_{i} 
   \nonumber \\
    & ~~ - \sum_{i,\nu}  \boldsymbol{S}_{i} \frac{\mathcal{J}_{\nu i}}{|\mu_{\nu}^{\mathrm{FM}}|} \delta\bm{\mu}_{S,\nu}
      + \sum_{i} \frac{\zeta_{i}\mu^d_{S,i} }{2\mu_{B}^{2}}  {\delta\bm{\mu}_{L,i}} \cdot
    \boldsymbol{S}_{i} 
    \nonumber \\
    & ~~ +  \sum_{i,\nu} \frac{\zeta_{\nu} |\mu_{\nu}^{\mathrm{FM}}| }{2\mu_{B}^{2}  |\sum
    _{n}\boldsymbol{S}_{0} \mathcal{J}_{\nu n} \boldsymbol{S}_{0}|}   {\delta\bm{\mu}_{L,\nu}} \mathcal{J}_{\nu i} \boldsymbol{S}_{i} \,  . 
    \label{eq:ham-renorm-with-field}
    \end{align}  

Here the electric field induced spin moments ($\delta\bm{\mu}_{S}$) couple through unrenormalized interatomic exchange  ($\mathcal{J}_{ij}$ in the $1^{\rm st}$ term) and intraatomic exchange ($J^{sd} = 0.5 $ eV  in the $2^{\rm nd}$ term). The induced orbital moments ($\delta\bm{\mu}_{L}$) enter via the spin-orbit coupling with strengths $\zeta_i=90.75$ meV and $\zeta_{\nu}=601.4$ meV, where $\nu$ denotes atoms in the Pt layers with proximity spin moment $\mu_{\nu}^{\rm FM}$ ($4^{\rm th}$ and $5^{\rm th}$ terms, respectively). Interface effects are well taken care of in the $3^{rd}$ and $5^{th}$ terms that lead to the renormalized quantities.  { All interaction parameters above were computed \textit{ab initio} and are taken from Ref.\ \cite{Devda2025}}.

The THz-field-induced spin ($\delta\bm{\mu}_{S,i}$) and orbital
($\delta\bm{\mu}_{L,i}$) moments entering the above Hamiltonian are
obtained from a layer-resolved magneto-electric susceptibility (MES)
tensor and depend linearly on the applied electric field,
\begin{align}
\delta \boldsymbol{\mu}_{S/L,i/\nu} & = \bm{\chi}^{S/L}_{i/\nu}(\theta,\phi)
\cdot \boldsymbol{E}(t) \, \label{eq:ind_mu_propto_E},
\end{align}
 {where $\bm{\chi}^{S/L}_{i/\nu}(\theta,\phi)$ is taken from Ref. \cite{Devda2025}, based on first principles calculation {s} within the Kubo linear-response formalism \cite{salemi2021}. The angles
$\theta$ and $\phi$ denote the magnetization direction, so that the
magnetization-dependent nature of the MES tensor is retained here.}

This MES tensor was originally evaluated for a DC electric
field. Applying it to a THz driving field is nevertheless justified,
because Guimar\~{a}es \textit{et al.} \cite{Guimaraes2020} showed that
the current-induced SOT effective fields remain close to their static
values over the whole range from gigahertz up to terahertz frequencies.
In other words, the induced moments follow the instantaneous field
rather than acquiring an appreciable frequency dependence of their own,
and the static susceptibility can be used to describe the response at THz
frequencies. We therefore evaluate Eq.~\eqref{eq:ind_mu_propto_E} at
each time step, using a broadband Gaussian THz pulse of $1$~ps duration,

\begin{equation}
\bm{E}(t)=\bm{E}_{0}
\exp\!\left[-\frac{\left(t-t_{\mathrm{c}}\right)^{2}}{\sigma^{2}}\right]
\sin\!\left[2\pi f\left(t-t_{\mathrm{d}}\right)\right],
\label{eq:Efield}
\end{equation}
 {where $\bm{E}_{0}$ sets the electric-field strength and $t_{\mathrm{d}}$
denotes the onset of the pulse, taken here as a delay of $50$~ps, so that
$t_{\mathrm{c}}=t_{\mathrm{d}}+1/(2f)$ is the centre of the Gaussian envelope.
The parameter $\sigma$ determines the temporal width of the envelope, while
$f$ is the carrier frequency. We use $\sigma=1.25\times10^{-13}$~s,
$f=1$~THz, corresponding to a single-cycle pulse of duration $1$~ps, and varied the range of $E_{0}$ values from $3 \times 10^8 $ to $3 \times 10^7$ ~V/m.}

The dynamics of  the Co spins is simulated using the stochastic Landau-Lifshitz equation (for finite temperature), 
\begin{align}
\frac{d\bm{S}_i}{dt}  & = -\frac{\gamma}{(1+\alpha^2)\mu_i}\bm{S}_i \times \left[ \left( \bm{H}^{\mathrm{eff}}_i + \bm{H}^\mathrm{SOT} \right) \right]  \nonumber \\
 & ~~ -\frac{\alpha\gamma}{(1+\alpha^2)\mu_i}\bm{S}_i \times  \bm{S}_i \times \left[ \,  \left( \bm{H}^{\mathrm{eff}}_i + \bm{H}^\mathrm{SOT} \right) \right] ,
 \label{SLLG_eq}
\end{align}

where $\gamma$ is the gyromagnetic ratio and $\alpha=0.01$ the Gilbert damping parameter. 
The total effective field arising from the Hamiltonian in Eq.\ \eqref{eq:ham-renorm-with-field_kurz} is given by

\begin{equation}
\bm{H}^{\mathrm{eff}}_i + \bm{H}^\mathrm{SOT}
=
-\frac{\partial (\mathcal{H}^{\rm eff} +\mathcal{H}^{\rm{SOT}})}{\partial \bm{S}_i}
+
\boldsymbol{\xi}_i(t) \, ,
\,\label{effective_field_withnoise}
\end{equation}
\noindent
where $\boldsymbol{\xi}_i(t)$ is a Gaussian white-noise field accounting for thermal fluctuations and satisfying the fluctuation-dissipation theorem. The SOT contribution enters the dynamics through $\bm{H}^{\mathrm{SOT}}$ which is directly proportional to the induced spin and orbital atomic moments, $\delta \mu_S$ and  $\delta \mu_L$. The SOT field possess both, a field-like  $\bm{H}^\mathrm{SOT}_\mathrm{FL} \propto (\boldsymbol{E} \times \hat{\boldsymbol{z}}) $ and an anti-damping-like character  $\bm{H}^\mathrm{SOT}_\mathrm{ADL} \propto - \boldsymbol{S}_{i} \times(\boldsymbol{E} \times \hat{\boldsymbol{z}}) $, both stemming from the induced moments. 

In this work, both the symmetric and asymmetric SOT configuration at the two NM/FM interfaces (as shown in Fig.\ \ref{fig:trilayer_schematics}(a)) are investigated, where the symmetric case emulates a Pt/Co/Pt {trilayer} with opposite polarities of {the} SOT at {the} two interfaces, while the asymmetric
case corresponds to a Pt/Co/Ta-type structure with similar polarities of SOT at {the} two interfaces, {since it is} known from previous work that Ta generates a SOT with polarity opposite to that of Pt \cite{Lee2021}. Note, however, that we do not use first-principles calculations for the Co/Ta interface, but we approximate it as another Co/Pt interface with reversed sign for the SOT.

\begin{figure*}
    \centering
    \includegraphics[scale=0.8]{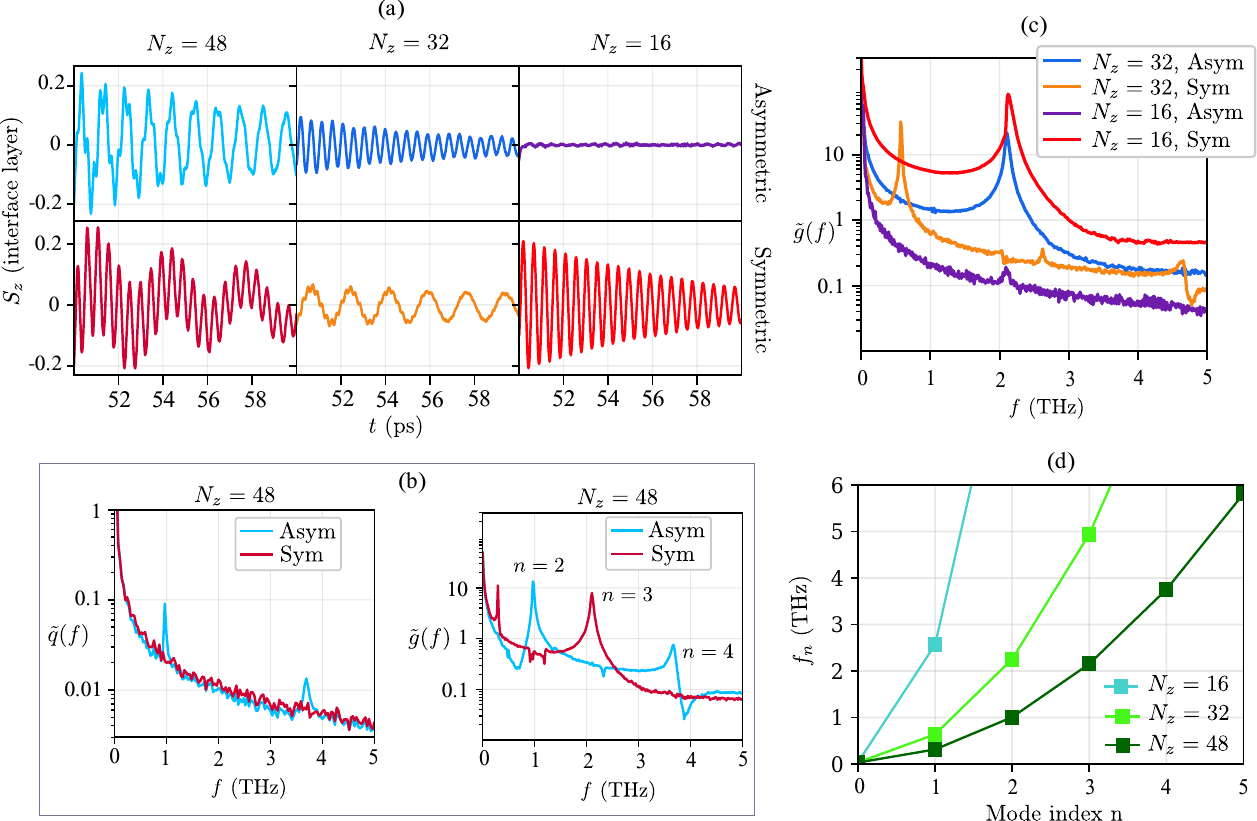}
\caption{(a) {Simulated} interface spin dynamics of {the} transverse spin component ($S_z$) after {THz} laser excitation, {for} different {Co} thicknesses $N_z$, {and for both, the symmetric and asymmetric configurations}. (b) Fourier transforms $\tilde{q}(f)$ and $\tilde{g}(f)$  of {the} transverse spin components 
{for the} symmetric and asymmetric SOT interfaces and $N_z=48$.
(c) $\tilde{g}(f)$ for thinner Co layers, $N_z=32$ and $N_z=16$.   {(d) Mode frequencies calculated analytically from the Kittel formula for
three Co layer thicknesses. The Kittel formula includes all frequencies which can occur and does not distinguish the symmetric or asymmetric configurations.}}
    \label{fig:Terahertz-excitation-peaks}
\end{figure*} 

\section{Results}

In the following, we simulate the THz spin dynamics of our sandwiched Co films for different thicknesses $t$ with  $N_z=48$ ($t\approx 9.9$  nm), $32$ ($t\approx 6.6$  nm) and $16$ ($\approx 3.3 $ nm in thickness) atomic layers in both, the {symmetric} (Sym-SOT) and {asymmetric} (Asym-SOT) configurations defined above. Upon application of an external static magnetic field $B_y = 400 \text{ mT}$, the equilibrium magnetization is stabilized in-plane along $\hat{y}$.  
The trilayer system is excited by a {THz}
electric-field pulse parallel to the equilibrium magnetization with peak amplitude $3 \times 10^{8}\,\mathrm{V/m}$,  {leading to field-like and anti-damping-like SOT-fields as shown in Fig.\ \ref{fig:trilayer_schematics}(b).}
The simulations were performed at a temperature of $T = 10\,\mathrm{K}$ in order to minimize thermal fluctuations and to ensure high spectral resolution in the Fourier analysis.

We start with a discussion of the dynamics of the transverse spin component $S_z$ at the Pt/Co interface layer as shown in Fig.\ \ref{fig:Terahertz-excitation-peaks}(a). Here for $N_z = 48$, both, symmetric and asymmetric configurations exhibit a superposition of distinct modes on picoseconds time scales, a first indication of the excitation of several quantized PSSW eigenmodes within the THz regime. The mixed-mode dynamics reduces to a single mode {upon} decreasing the {Co} thickness {to}  $N_z=32$ and $16$. 

To extract the modes of these spin-precession dynamics, we {define} the spectral power density in terms of two quantities obtained from the Fourier transform of {the} average magnetization as, 
\begin{equation}
    \tilde{q}(f) = \left\langle  \left| \sum_{n=0}^{N-1} \tilde{m}(t_n)\, e^{-i\,2\pi f\, t_n} \right|^2 \right\rangle ,
\end{equation}
and 
\begin{equation}
    \tilde{g}(f) = \left\langle  \left| \sum_{n=0}^{N-1} \tilde{n}(t_n)\, e^{-i\,2\pi f\, t_n} \right|^2 \right\rangle ,
\end{equation}
where, $\left<...\right>$ represents the thermal ensemble average, discrete times $t_n = n\,\Delta t$ with $n = 0, \ldots, N-1$ correspond to the sampled magnetization trajectory, and $\Delta t$ is the sampling interval. Here  $ \tilde{m}$ and $\tilde{n}$ are complex functions of the average transversal magnetization components, given as 
\begin{equation*}
\tilde{m}(t_n) = \, \overline{m_z}\, (t_n) +i \, \overline{m_x}\, (t_n),~~ \tilde{n}(t_n) = \, \overline{m_z^2}\, (t_n) +i \, \overline{m_x^2}\, (t_n).
\end{equation*}
 {Since $\tilde{m}(t)$ is linear in the transverse components, $\tilde{q}(f)$ shows peaks at the fundamental eigenfrequencies
$f_i=\omega_i/2\pi$ of the excited PSSW modes.  However, odd modes would vanish in the spatially averaged transverse magnetization, particularly for
symmetric interfaces, and are therefore not resolved by $\tilde{q}(f)$. The quadratic signal $\tilde{n}(t)$ solves this problem and a mode at $\omega_i$ contributes
a peak in  $\tilde{g}(f)$ at frequency $2\omega_i$.
For direct
comparison, $\tilde{g}$ is plotted against $f/2$, so that peaks in
$\tilde{q}(f)$ and $\tilde{g}(f)$ arising from the same eigenmode fall on a common frequency axis.}

As shown in Fig.~\ref{fig:Terahertz-excitation-peaks}(b) for $N_{z}=48$, the quantity $ \tilde{q}(f) $ clearly reveals the excitation of higher-order perpendicular standing spin-wave modes for the asymmetric SOT configuration, with frequencies extending into the high {THz} regime. In contrast, for the symmetric SOT configuration, no pronounced excitation peaks are observed. However, when performing the Fourier transform of the squared transverse spin component, $\tilde{g}(f)$, distinct excitation peaks emerge for the symmetric configuration as well. This behavior originates from the symmetry of the interfacial torques. In the symmetric configuration, SOTs generated at the two interfaces possess opposite polarities. When averaged across the film thickness, these opposing torques largely cancel each other, thereby suppressing any net excitation. Consequently, such modes cannot be detected in conventional pump–probe experiments, whereas interface-sensitive measurements may still reveal their presence.
In contrast, the asymmetric SOT configuration produces interfacial torques of identical polarity, resulting in constructive excitation of specific PSSW modes.

The PSSW mode frequencies can be identified using {the} Kittel {formula} {with {dependent} field contributions. The frequency of the \(n\)-th standing spin-wave mode
can be written as}

\begin{equation}
f_n =
\frac{\gamma}{2\pi}
\sqrt{
\left(\bar{H}^{\mathrm{eff}}_y + D_{\mathrm{eff}} k_z^2\right)
\left(\bar{H}^{\mathrm{eff}}_z + D_{\mathrm{eff}} k_z^2\right)
},
\label{eq:kittel_pssw}
\end{equation}
  {where the wavevectors $k_z$ are given by
\begin{equation}
k_z = \frac{2\pi n}{N_z a_z},
\quad
n = 0,1,2,\dots,\frac{N_z}{2},
\end{equation}
with $n$ denoting the mode index. The averaged effective fields are obtained as,
\begin{equation}
\bar{H}^{\mathrm{eff}}_{\alpha}
=
\frac{1}{N_z}
\sum_{i=1}^{N_z}
H^{\mathrm{eff}}_{\alpha,i},
\qquad
\alpha\in\{y,z\}.
\end{equation}
Here, for each layer \(i\), the local effective fields are 
\begin{align}
H^{\mathrm{eff}}_{y,i}
&=
H_y
+
\frac{K_{xx,i}-K_{yy,i}}{\mu_i}
+
\frac{\mathcal{J}_{xx,i}-\mathcal{J}_{yy,i}}{\mu_i},
\label{eq:Hyeff_layer}
\\[4pt]
H^{\mathrm{eff}}_{z,i}
&=
H_y
+
\frac{K_{xx,i}-K_{zz,i}}{\mu_i}
+
\frac{\mathcal{J}_{xx,i}-\mathcal{J}_{zz,i}}{\mu_i},
\label{eq:Hzeff_layer}
\end{align}
where \(H_y\) denotes the applied Zeeman-field contribution, \(K_{\alpha
\alpha,i}\) are the diagonal components of the on-site anisotropy tensor in
layer \(i\), and \(\mathcal{J}_{\alpha\alpha,i}\) is the total diagonal
exchange contribution from all {interaction} bonds within that layer,
\begin{equation}
\mathcal{J}_{\alpha\alpha,i}
=
\sum_{j\in\mathrm{nbrs}(i)}
J^{ij}_{\alpha\alpha},
\qquad
\text{for }\alpha\in\{x,y,z\}.
\end{equation}
This construction includes both the magnetocrystalline anisotropy and the
anisotropy of the exchange tensor in the local restoring fields. 
The {effective exchange stiffness} {$D_{\rm eff}$} in Eq.\  \eqref{eq:kittel_pssw} is defined as
\begin{equation}
D_{\mathrm{eff}} =
\frac{1}{N_z}
\sum_{i=1}^{N_z}
\frac{2 J_{\perp,i} a_z^2}{\mu_i} \, ,
\end{equation}
where $J_{\perp,i}$ represents interlayer exchange interactions for each Co layer \(i\)}.

The analytical mode calculation (see Fig.\ \ref{fig:Terahertz-excitation-peaks}(d)) confirms that the asymmetric SOT configuration selectively excites even modes, specifically $n=2$ at $f \approx 1\,\mathrm{THz}$ and $n=4$ at $f \approx 3.7\,\mathrm{THz}$. In contrast, the symmetric configuration predominantly excites odd modes, namely $n=1$ at $f \approx 0.35\,\mathrm{THz}$ and $n=3$ at $f \approx 2.1\,\mathrm{THz}$. These mode assignments are consistent with the  {Fourier transform} spectra obtained from the squared spin components and are in qualitative agreement with the experimental work of {Salikhov} \textit{et al.}\ \cite{Salikhov2023}, although quantitative differences in the absolute frequencies arise from material-specific parameters in the present study. Upon reducing the Co thickness, a systematic modification of the PSSW spectrum is observed. For $N_z = 32$, under the same electric-field amplitude, the asymmetric SOT configuration excites the $n=2$ mode at approximately $2\,\mathrm{THz}$, while the symmetric configuration excites the $n=1$ mode at $0.5\,\mathrm{THz}$ and the $n=3$ mode reaching $4.7\,\mathrm{THz}$. 
Further reducing the {Co} thickness to $N_z = 16$ significantly increases the mode spacing, pushing the energy gap into the higher THz regime. In this ultrathin limit, the asymmetric SOT at given electric field magnitude becomes insufficient to excite higher-order modes, whereas the symmetric configuration still excites the lowest odd mode at approximately $2\,\mathrm{THz}$. 
This thickness dependence directly reflects the quantization condition and the increasing exchange energy cost associated with higher-order standing modes in thinner films. 

\begin{figure}[t]
    \centering
    \includegraphics[scale=0.95]{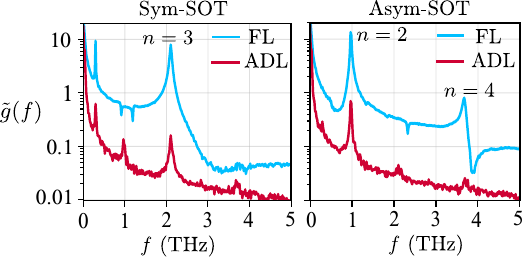}
    \caption{{Fourier transform} of the time-dependent spin component, simulated independently in the presence of either $M$-even (field-like) or $M$-odd (antidamping-like) SOT for $N_z=48$ at $E=3 \times 10^8$ V/m.}
    \label{fig:excitation with_even_odd_SOT}
\end{figure}
To address the question, which of {the} microscopic torque components drives each excitation, we separately activate the field-like and anti-damping-like torques as our microscopic SOT framework enables a detailed separation of {these} torque components, previously identified as exhibiting magnetization-even ($M$-even) and magnetization-odd ($M$-odd) symmetry, respectively \cite{Devda2025,salemi2021}. By simulating the asymmetric SOT configuration with $M$-even and $M$-odd SOT treated independently, a clear distinction in their excitation efficiency emerges, {as shown in} Fig.\ \ref{fig:excitation with_even_odd_SOT}. The excitation peaks corresponding to the $n=1$ and $n=3$ modes in the symmetric configuration, as well as the $n=2$ and $n=4$ modes in the asymmetric configuration, are prominently observed for the $M$-even (field-like) SOT component, whereas the $M$-odd (anti-damping-like) SOT alone is comparatively inefficient in exciting these modes. 

 {
This behavior is consistent with the experimental picture reported by Salikhov \textit{et al.}\ \cite{Salikhov2023}, where the field-like torque was identified as the dominant driving mechanism for the excitation of these THz PSSW modes and where even modes become accessible only in case of asymmetric interfaces. Our simulations reproduce both features from a purely microscopic starting point, without imposing any phenomenological interfacial excitation profile.}

\begin{figure}[t]
    \centering
    \includegraphics[scale=0.9]{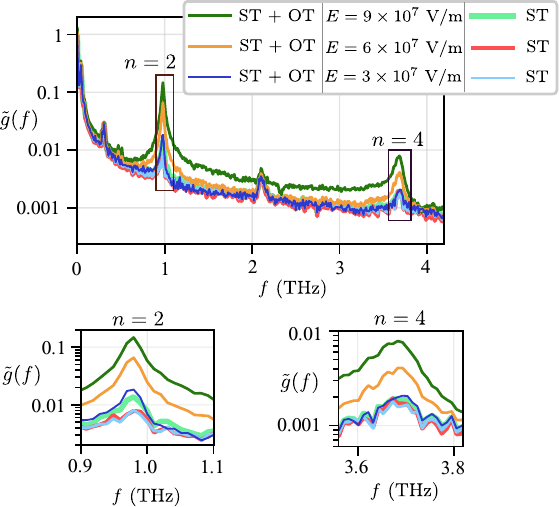}
    \caption{Mode excitation in {the} asymmetric-SOT configuration {computed}  in {the} presence {or} absence of {the} orbital torque (OT) in addition to {the} spin torque (ST) under laser pulse excitation with different electric field magnitudes. {Insets show zooms into the peaks at $n=2$ and $n=4$.} }
    \label{fig:separating_spin_orbital}
\end{figure}

Next, to answer the question, what is the most prominent contribution of the field-like SOT, we continue with separating orbital and spin torque components from the SOT.
To this end, we separated orbital and spin torque contributions by turning on and off the orbital contribution terms in the SOT Hamiltonian, while varying the experimentally relevant electric-field amplitude. Since {our} primary objective is to analyze the role of {the} orbital torque (OT) {versus the spin torque} in the excitation of THz PSSW modes, this decomposition allows a direct comparison of their respective efficiencies.  The  {Fourier transform} spectra of {the magnetization component for} the asymmetric SOT {configuration, shown in} Fig.~\ref{fig:separating_spin_orbital}, reveals that in the presence of both, orbital and spin torques, the even modes exhibit significantly enhanced amplitudes at $E = 90\,\mathrm{mV/nm}$ and remain clearly visible even at a reduced field strength of $E = 30\,\mathrm{mV/nm}$. In contrast, in the absence of {the} orbital torque, even a high electric-field amplitude of $90\,\mathrm{mV/nm}$ is insufficient to efficiently excite the lower even-order mode ($n=2$).

\section{Discussion and Conclusions}

 { {Our} atomistic spin-model simulations establish that the excitation of exchange-dominated PSSW modes is governed predominantly by field-like orbital torques. In the $9.9$~nm Co film, modes up to the fourth order are excited, reaching $f = 3.7$~THz, while in the $6.6$~nm film the second-order mode appears near $2$~THz. In comparison, the damping-like and spin-torque contributions produce substantially weaker spectral responses. This hierarchy is observed for both even and odd modes, both are strongly enhanced only when the field-like orbital torque is included.}

 {Our  findings provide a microscopic interpretation of the experimentally observed THz PSSW excitation reported by Salikhov \textit{et al.}~\cite{Salikhov2023}. Their measurements showed that the field-like SOT drives THz magnon excitation, but the reason for its dominance over the damping-like torque and the relative role of spin and orbital contributions, remained unresolved, since optical and transport probes access only the total torque response~\cite{Ding2020,Hayashi2023,Lee2021}. Our decomposition of the total torque into spin and orbital as well as field-like and damping-like contributions 
 {identifies} the microscopic origin of the observed PSSW excitation.}

The dominance of the field-like orbital torque can be traced to its strong localization within a few atomic layers of the Co/NM interfaces, as predicted by first-principles calculations~\cite{Devda2025,Alikhah2026}. This localization generates a highly non-uniform spin excitation profile across the ferromagnetic layer thickness. That profile provides precisely the symmetry breaking required to couple efficiently to standing spin-wave modes of both lower and higher order. Its efficiency is reinforced by the weak dependence of the orbital torque on the magnetization direction~\cite{Devda2025}. As the magnetization is driven far from equilibrium during the THz pulse, conventional torque components that track the instantaneous magnetization~\cite{Mahfouzi2018,Belashchenko2019,Chuang2020} are dynamically modified. The field-like orbital torque, in contrast, retains an almost constant amplitude and provides a more persistent interfacial driving field.

 {Sensitivity to spatially confined angular-momentum transfer has also been reported for interfacial spin-transfer torques generated by superdiffusive spin currents~\cite{Razdolski2017,Ulrichs2018,Brandt2021,Weissenhofer2023}. In those cases the angular momentum is carried by hot electrons {injected into an FM} across an NM spacer. Here,  {conversely}, it originates from the interfacial orbital angular momentum generated by the driving field itself. This distinction is relevant because orbital currents exhibit relaxation lengths and interfacial conversion efficiencies substantially different from their spin counterparts~\cite{Seifert2023,Hayashi2023,Guan2026}. The direct-field route also avoids the spacer and hot-electron transport of the earlier schemes, offering a more efficient path to exchange-dominated THz magnons. The persistence of the effect across symmetric and asymmetric configurations identifies interfacial orbital-angular-momentum transfer, rather than spin torque alone, as the primary excitation channel. Engineering its magnitude, localization, and relative sign through material and interface design~\cite{Dongwook2020,Ding2020} provides a promising route toward selective and efficient THz magnon excitation.}

\begin{acknowledgments}
This work has been supported by the the German Research Foundation (Deutsche Forschungsgemeinschaft) through CRC/TRR 227 ``Ultrafast Spin Dynamics'' (project MF, project-ID: 328545488)) and CRC 1432 (project B02, project-ID: 425217212), through the Knut and Alice Wallenberg Foundation (Grants No.\ 2022.0079 and No.\ 2023.0336), and by the EIC Pathfinder OPEN Grant No.\ 101129641 (OBELIX). Computational resources were provided by the core facility SCCKN in Konstanz.
\end{acknowledgments}

\bibliography{THz_SOT}

\end{document}